\documentclass{pasj00}

\newcommand{\pasa}{PASA} 

\begin{document}
\SetRunningHead{Niino, Totani \& Okumura}{Optical Follow Up of FRBs}

\title{Unveiling the Origin of Fast Radio Bursts by Optical Follow Up Observations} 


\author{
Yuu \textsc{Niino}\altaffilmark{1}
Tomonori \textsc{Totani}\altaffilmark{2,3}
and
Jun E. \textsc{Okumura}\altaffilmark{2,4}}
\altaffiltext{1}{Division of Optical and IR Astronomy, 
National Astronomical Observatory of Japan, 2-21-1 Osawa, Mitaka, Tokyo}
\email{yuu.niino@nao.ac.jp}
\altaffiltext{2}{Department of Astronomy, School of Science, The University of Tokyo, 7-3-1 Hongo, Bunkyo-ku, Tokyo}
\altaffiltext{3}{Research Center for the Early Universe, School of Science, The University of Tokyo, 7-3-1 Hongo, Bunkyo-ku, Tokyo}
\altaffiltext{4}{Department of Astronomy, Kyoto University, Kitashirakawa-Oiwakecho, Sakyo-ku, Kyoto}



\maketitle

\begin{abstract}
We discuss how we can detect and identify counterparts of fast radio bursts (FRBs) 
in future optical follow up observations of FRBs if real-time alert of FRBs becomes available. 
We consider kilonovae as candidates of FRB optical counterparts, 
as expected in the case that FRBs originate 
from mergers of double neutron star binaries. 
Although theoretical predictions on luminosities of kilonovae are still highly uncertain, 
recent models suggest that kilonovae can be detected 
at redshifts up to z $\sim$ 0.3 within the range of the uncertainties. 
We expect $\sim$ 1--5 unrelated supernovae (SNe)
down to a similar variability magnitude in 5 days interval 
within the typical error radius of a FRB. 
We show that, however, a kilonova can be distinguished from
these SNe by its rapid decay and/or color evolution,
making it possible to verify the existence of a kilonova associated with a FRB. 
We also discuss the case that SNe Ia are FRB optical counterparts, 
as it might be if FRBs originate from double white dwarf binaries. 
Verification of this scenario is also possible, 
since the chance probability of finding a SNe Ia having
consistent explosion time with that of a FRB within the FRB 
error region is small (typically $\lesssim$ 0.01). 
\end{abstract}

\section{Introduction}

Fast Radio Bursts (FRBs) are transient events recently discovered 
at $\sim$ GHz frequency \citep{Lorimer:2007a, Keane:2012a, Thornton:2013a, Spitler:2014a}. 
FRBs have typical durations of several milliseconds. 
Their large dispersion measures (DMs) suggest 
that FRBs are at cosmological distances corresponding to $z\sim$ 0.3--1. 

Some models of the origin of FRBs were proposed recently, 
including merger of double neutron star binaries (NS-NS, \cite{Totani:2013a}), 
merger of double white dwarf binaries (WD-WD, \cite{Kashiyama:2013a}), 
hyperflares of magnetars \citep{Popov:2013a}, 
collapse of rotating super-massive neutron stars 
to black holes \citep{Falcke:2014a, Zhang:2014a}, 
galactic flaring stars (\cite{Loeb:2014a}, 
but see also \cite{Tuntsov:2014a}; \cite{Dennison:2014a}), 
galactic exotic compact objects \citep{Keane:2012a, Bannister:2014a}, 
and Perytons \citep{Burke-Spolaor:2011a, Kulkarni:2014a}. 

The current localization errors of the FRBs are too large 
to identify host galaxies (or galactic progenitors) of FRBs. 
Discovery of FRB counterparts in other wavelengths 
is important to unveil the origin of FRBs. 
The currently known FRBs have been found in post analyses of pulsar survey data, 
and hence no follow up observation could be performed at the time of the bursts. 
Real-time alert and triggered multi-wavelength follow up observations 
are desirable to improve our understanding on FRBs. 

Some of the proposed FRB models predict 
the existence of counterparts in other wavelengths (e.g. \cite{Yi:2014a}). 
In the context of future FRB follow up observations in optical/near-infrared (NIR) wavelengths, 
the cases that a FRB results from a NS-NS or WD-WD merger, are particularly interesting. 
It is expected that NS-NS mergers cause a couple of transient events 
other than FRBs, namely short gamma-ray bursts (GRBs) and kilonovae. 
A kilonova is a transient event in optical/NIR powered by decay of radioactive elements 
formed via $r$-process nucleosynthesis in ejecta from a NS-NS merger. 

The existence of kilonovae has been theoretically predicted \citep{Li:1998a}, 
and a kilonova candidate was recently discovered in follow up observations 
of short GRB 130603B \citep{Tanvir:2013a, Berger:2013a}. 
Kilonova emission lasts for a duration of days to weeks. 
Furthermore, kilonova emission is isotropic, 
and hence we would be able to detect it 
from every NS-NS merger that is close enough, 
unlike short GRBs that are most likely beamed. 
In this letter, we investigate how we can detect and distinguish kilonovae 
in future optical follow up observations of FRBs 
in the case that FRBs originate from NS-NS mergers. 

We also briefly discuss the case that FRBs originate from WD-WD mergers. 
WD-WD mergers are promising candidates of the origin of type Ia supernovae (SNe Ia), 
although there are also other promising candidates 
(see \cite{Maoz:2013a} for a recent review). 
If a SN Ia is associated with a FRB, 
it will be a good target of optical follow up observations \citep{Kashiyama:2013a}. 

Throughout this letter, we assume the fiducial cosmology 
with $\Omega_{\Lambda}=0.7$, $\Omega_{m}=0.3$, and $H_0=$ 70 km s$^{-1}$ Mpc$^{-1}$.
The magnitudes are given in the AB system. 

\section{Detectability of Kilonovae Associated with FRBs}
\label{sec:models}

Recently several authors have investigated spectra and light curves 
of kilonovae taking opacity of $r$-process elements into account 
\citep{Barnes:2013a, Tanaka:2013a, Grossman:2014a}. 
Although the models of kilonovae are still uncertain, 
it is broadly agreed that the optical light curve rapidly decays within a few days after the explosion, 
while the NIR light curve gradually rises and peaks several days later. 

\begin{figure*}
 \begin{center}
  \includegraphics[width=16cm]{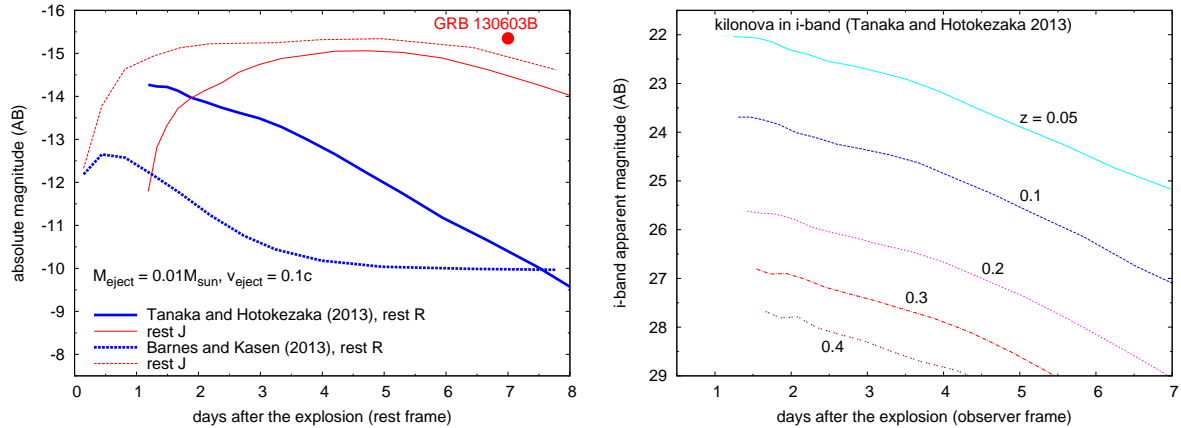} 
 \end{center}
\caption{
Left panel: rest frame light curves of kilonovae 
in absolute magnitude predicted by the models 
of \authorcite{Tanaka:2013a} (\yearcite{Tanaka:2013a}, TH13) and \citet{Barnes:2013a}. 
The thick and thin lines represent 
the light curves in rest frame $R$-band and $J$-band, respectively. 
The circle at the top right indicates observing time and luminosity 
of the kilonova candidate associated with GRB 130603B ($z=0.356$) 
in HST F160W filter which roughly corresponds to rest frame $J$-band. 
Right panel: apparent light curves in observer frame $i$-band 
predicted by the TH13 model at different redshifts. 
}\label{fig:modelLC}
\end{figure*}

In the left panel of figure~\ref{fig:modelLC}, we show kilonova light curves 
predicted by \citet{Tanaka:2013a} and \citet{Barnes:2013a} 
using fiducial parameter sets in their studies. 
The models are broadly consistent in their late time NIR light curves, 
while the light curves in the first couple of days after the explosion 
and/or in optical wavelengths are different from each other. 
Besides the model uncertainties, the luminosity of a kilonova largely depends 
on some parameters such as mass and velocity of the ejecta from the merging NS-NS binary. 
It is possible that kilonovae span a wide range of peak luminosities, 
although their luminosity function is not known. 
In the following discussion, we use the fiducial model of 
\authorcite{Tanaka:2013a} (\yearcite{Tanaka:2013a}, hereafter TH13 model)
to make a rough estimate of the feasibility of optical follow up observations. 

The currently known FRBs are discovered 
by the multibeam receiver on the 64-m Parkes radio telescope \citep{Staveley-Smith:1996a}
and the Arecibo L-band Feed Array on the 305-m Arecibo telescope \citep{Cordes:2006a}. 
With these radio telescopes, 
the typical localization error radius of a FRB is several arcmin, 
which can be covered by wide-field cameras of 8m-class telescopes 
such as Suprime Cam/Hyper Suprime Cam (SCam/HSC) on Subaru. 

The kilonova models predict that the light curves in shorter wavelength range decay earlier, 
and hence the optical follow up should be performed in red bands. 
With 8-m class telescopes, an $i$-band image 
with limiting magnitude $\sim$ 27.5 (S/N = 5)
can be achieved with 3--4 hours of exposure. 
In the right panel of figure~\ref{fig:modelLC}, 
we show the observer frame $i$-band light curves of kilonovae 
at various redshifts predicted by the TH13 model. 
We may detect kilonovae at redshifts up to $z \sim 0.3$ 
if we perform an imaging down to $i\sim27.5$ within $\sim$ 3 days after FRBs. 
Hence FRBs with DM smaller than that corresponding to $z \sim 0.3$ 
will be good targets of the optical follow up observations. 
The expected peak NIR magnitude of a kilonova at $z \sim 0.3$ is $H\sim26$. 
Hence it would be difficult to detect a kilonova 
at a cosmological distance with a ground based NIR telescope. 

Assuming a constant FRB rate 
and the maximum observable redshift of FRBs to be $z=1$, 
the FRB rate reported in \citet{Thornton:2013a} 
becomes $2.4^{+1.4}_{-1.2}\times10^4$ yr$^{-1}$ Gpc$^{-3}$. 
Then the event rates of $z < 0.3$ FRBs in the field of views 
of the Parkes multibeam receiver and the Arecibo L-band Feed Array 
\citep{Staveley-Smith:1996a, Spitler:2014a}
are $\sim 2.3$ yr$^{-1}$ and $\sim 0.45$ yr$^{-1}$, respectively. 

We note that \citet{Hassall:2013a} showed that 
FRB rate estimates change by an order of magnitude depending 
on the treatment of the scattering of radio pulses as they propagate through a plasma. 
It is also likely that the FRB rate density depends 
on redshift as discussed in \citet{Totani:2013a}. 
The observed redshift evolution of the SN Ia rate density is
expressed as $\propto (1+z)^2$ \citep{Okumura:2014a}. 
We expect that the redshift dependence of 
the rate density of NS-NS mergers would be similar to that of SN Ia, 
because the delay time distribution of NS-NS mergers is expected 
to be $\propto t^{-1}$ \citep{Totani:1997a, Dominik:2013a} 
and the same delay time distribution is also suggested 
for SNe Ia \citep{Totani:2008a, Maoz:2013a}. 

\section{Estimation of Contaminating SN Rate} 

When we search for an optical counterpart of a FRB, 
it is possible that physically unrelated SNe 
coincide within the positional error by chance. 
To investigate the population of SNe 
that contaminate the FRB follow up observations, 
we randomly generate a mock catalog of SNe in the field 
considering SN types Ia, Ibc, IIL, IIP, and IIn at various redshifts. 

Redshifts of the mock SNe are sampled 
at intervals of $\Delta z = 0.05$ up to $z = 2$ ($z = 4$ for IIn) 
where the SNe become undetectable at the peak times 
by currently available facilities ($i > 27.5$). 
At each sampled redshift, we randomly generate 1000 sets 
of light curves in observer frame $i$ and $z$-band for each of the SN types, 
considering peak time luminosity distribution of each SN type. 
When generating SN Ia light curves, we follow the method of \citet{Barbary:2012a} 
in which luminosity of a SN Ia correlates with its stretch and color. 
For core-collapse SNe (Ibc, IIL, IIP, and IIn), we follow the method of \citet{Dahlen:2012a}. 
Dust extinction in SN host galaxies are included in these models. 
Each light curve follows the templates of SN spectral evolutions 
provided by \citet{Hsiao:2007a} and \citet{Nugent:2002a}. 

For each of the generated light curves, 
we consider various timings of the observation 
ranging -30--1000 days relative to the peak time of each SN sampled at intervals of 1 day. 
For each of the sampled timings, 
we obtain apparent $i$ and $z$ magnitudes and their later evolution. 
To simulate the SN population which we would find in the field, 
we integrate the mock SN sample over redshifts 
weighted with cosmic SN rate history of each SN type. 
We assume SN Ia rate history of \citet{Okumura:2014a}, 
and core-collapse SN rates proportional 
to the cosmic star formation rate history \citep{Behroozi:2013a} 
normalized to core-collapse SN rates at $z<1$ \citep{Dahlen:2012a}. 

We consider a case that we take a residual of 2 images 
taken with a certain time interval to find transient events. 
The interval of several days would be sufficient to find a kilonova due to their rapid decay. 
We note that, when a transient is embedded 
on a background object such as its host galaxy and we cannot separate them, 
the flux information we can obtain is limited to $f_{\rm \nu,1st}-f_{\rm \nu,2nd}$, 
where $f_{\rm \nu,1st}$ and $f_{\rm \nu,2nd}$ 
are the fluxes of the transient in the first and the second image. 
In this case, we need another (third) image with a sufficient time interval
as a reference frame that does not include light from the transient,
to measure the total (i.e., non-residual) flux 
of the transient, $f_{\rm \nu,1st}$ and $f_{\rm \nu,2nd}$, separately. 

\begin{figure}
 \begin{center}
  \includegraphics[width=7cm]{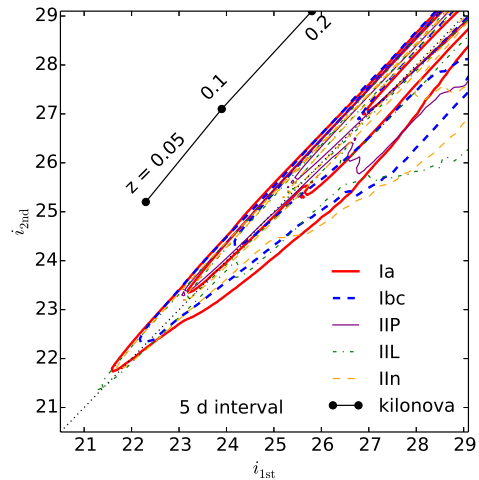} 
 \end{center}
\caption{
Distribution of SNe on the $i_{\rm 1st}$ vs. $i_{\rm 2nd}$ plane with the interval of 5 days. 
The contours indicate number density of the SNe on this plane of 1, 10, and 100 mag$^{-2}$deg$^{-2}$. 
The 5 SN types considered in our SN model (Ia, Ibc, IIL, IIP, and IIn) are plotted separately. 
Kilonvae at different redshifts are plotted together 
with the first observation at 2 days after the explosion. 
The dotted line indidates $i_{\rm 1st} = i_{\rm 2nd}$. 
}\label{fig:MMmap}
\end{figure}

In figure~\ref{fig:MMmap}, we show distribution of $i$-band magnitudes of each SN type 
in the first and second image ($i_{\rm 1st}$ and $i_{\rm 2nd}$) taken with the intervals of 5 days. 
SNe are distributed along the $i_{\rm 1st} = i_{\rm 2nd}$ line,  
although a few SNe are in rapidly rising phase ($i_{\rm 1st} > i_{\rm 2nd}$). 
When the interval is shorter the scatter of the SN distribution 
around the $i_{\rm 1st} = i_{\rm 2nd}$ line is smaller 
($< 0.1$ mag in the case of the 1 day interval). 

\section{Distinguishing a FRB and SNe} 
\label{set:distinguish}

\begin{figure}
 \begin{center}
  \includegraphics[width=8cm]{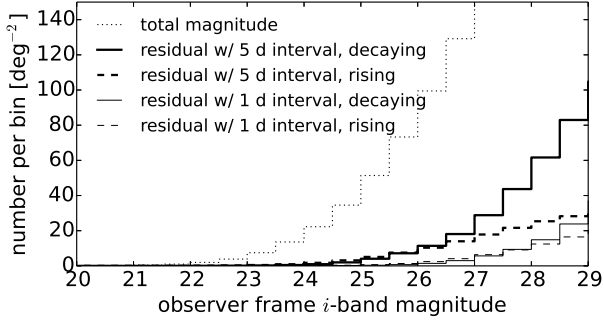} 
 \end{center}
\caption{
Differential distributions of $i$-band magnitudes of contaminating SNe (all types). 
The dotted histogram represents the total magnitudes, 
while the other histograms represent those in the residual images 
with the intervals of 5 days and 1 day (thick and thin histograms, respectively). 
For the residual magnitudes, the solid and dashed lines 
represent the decaying and rising SNe, respectively. 
}\label{fig:ires_SN}
\end{figure}

In figure~\ref{fig:ires_SN}, we show distributions 
of total $i$-band magnitudes of SNe and those in the residual images 
with the intervals of 5 days and 1 day predicted by the mock SN catalog. 
The magnitude in the residual image ($m_{\rm res}$) can be expressed 
as $m_{\rm res} = -2.5\ {\rm log}_{10}|f_{\rm \nu,1st}-f_{\rm \nu,2nd}|-48.6$, 
where $f_{\rm \nu,1st}$ and $f_{\rm \nu,2nd}$ are in a unit of erg s$^{-1}$cm$^{-2}$Hz$^{-1}$. 
The $i_{\rm res}$ distribution of rising and decaying SNe 
are plotted separately in figure~\ref{fig:ires_SN}. 
When we search for a kilonova in the field, 
we consider only decaying SNe as contaminants. 
Although $i_{\rm res}$ of SNe 
are significantly fainter than their total magnitudes, 
we would detect some contaminants in a deep survey.  
In a survey down to the depth of $i_{\rm res} = 27.5$, 
the number density of contaminants is $73.5$ deg$^{-2}$ 
with the interval of 5 days (i.e. 1.6 SNe within an error radius of 5 arcmin). 
The number density of contaminants is smaller with the 1 day interval. 
However, kilonovae decay only by $\sim$ 50\% 
of their luminosity within the 1 day interval, 
and $i_{\rm res}$ of kilonovae would also become significantly fainter 
than the total magnitudes making kilonova detection more difficult. 

\begin{figure}
 \begin{center}
  \includegraphics[width=8cm]{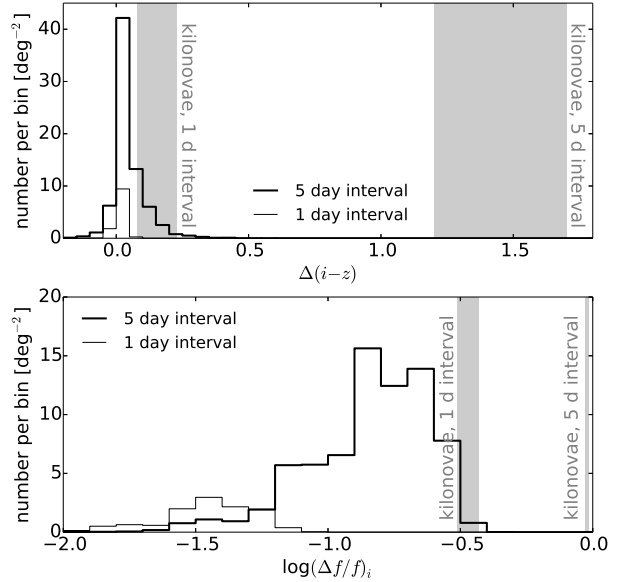} 
 \end{center}
\caption
{
Distributions of the color evolution of SNe [$\Delta(i-z)$, top panel] 
and the fractional variability flux in $i$-band ($\Delta f/f$, bottom panel), 
with the time intervals of 5 days and 1 day. 
Only decaying SNe with $i_{\rm res} < 27.5$ are considered. 
The lines are the same as those in figure~\ref{fig:ires_SN}. 
The shaded regions indicate $\Delta(i-z)$ and $\Delta f/f$ 
of kilonovae at $z \leq 0.3$ with the first observation 
at 2 days after the explosion. 
}\label{fig:distinguish}
\end{figure}

\citet{Tanaka:2013a} showed that kilonovae have redder color than SNe throughout their evolution. 
However, it should be noted that SNe may have similar apparent color to kilonovae, 
when reddened by dust or cosmologically redshifted.  
One clue to distinguish kilonovae from SNe is the color evolution. 
Color of a kilonova changes by $\Delta(i-z) \gtrsim 1$ 
in the first several days after the explosion, 
while SN colors do not significantly change in such a timescale (the top panel of figure~\ref{fig:distinguish}). 
Here we consider a case of a FRB follow-up starting at 2 days after the explosion, 
followed by the second optical observation with an interval of 1 or 5 days from the first. 
The large color evolution of $\Delta(i-z) \gtrsim 1$ 
can be realized over photometric errors when kilonovae are detected with S/N $> 5$, 
although detection in the second image may be available 
only for kilonovae at $z \lesssim 0.1$ and/or brighter events than considered here. 

Another clue to distinguish kilonovae from SNe 
is the rapid decay of kilonovae in optical wavelengths as shown in figure~\ref{fig:MMmap}. 
In the bottom panel figure~\ref{fig:distinguish}, we show distribution of the fractional variability, 
 $\Delta f/f = (f_{\rm \nu,1st}-f_{\rm \nu,2nd})/f_{\rm \nu,1st}$, 
in $i$-band for decaying SNe with $i_{\rm res} < 27.5$. 
SNe have $\Delta f/f \lesssim 0.3$ (0.1), while a kilonova reduces its flux 
by $\Delta f/f \gtrsim$ 0.9 (0.3) during the 5 days (1 day) interval. 
The fractional variability can be used to distinguish a kilonova from SNe, 
even when the kilonova is not detected in the second image. 
If we perform $i$-band imaging of a kilonova at $z = 0.3$ 
at 2 days and 7 days after the explosion with the limiting magnitude of $i = 27.5$, 
the magnitude and the limit which we obtain in the two images 
would be $i_{\rm 1st} = 27.0$ and $i_{\rm 2nd} > 27.5$, 
respectively (the right panel of figure~\ref{fig:modelLC}). 
In this case, we obtain a limit of $\Delta f/f > 0.37$ 
which is sufficient to distinguish the kilonova from SNe. 
Thus it is possible to distinguish kilonovae with $i$-band magnitudes 
down to $i_{\rm 1st} \sim 27$ (i.e 0.5 mag above the limit). 

It is also possible that active galactic nuclei (AGNs) and/or 
galactic variables contaminate the follow up observations. 
However, AGNs with variability timescale $\lesssim 10$ days are rare 
(e.g. \cite{Totani:2005a}; \cite{Morokuma:2008a}). 
Furthermore, AGNs would always appear at the center of their host galaxies, 
while kilonovae have wide variety of offsets from the center 
as suggested from observations of short GRBs (1--50 kpc, \cite{Fong:2010a}). 
The number density of galactic variables are also smaller 
than that of SNe at high galactic latitudes ($\sim$ 1/4, \cite{Morokuma:2008a}), 
We also note that galactic variables typically 
have $\Delta f/f \lesssim 0.3$ in $i$-band (e.g. \cite{Ivezic:2000a}), 
and would not be associated with a candidate host galaxy in most cases. 
Thus it would not be difficult to distinguish a kilonova from AGNs and galactic valuables. 

\section{Discussion}

We have investigated the detectability of FRB optical counterparts 
in the scenario that FRBs originate from NS-NS mergers as proposed by \citet{Totani:2013a}. 
NS-NS mergers may accompany radioactive emissions called kilonovae. 
Although the recent models of kilonovae are still highly uncertain, 
they suggest that a kilonova may be detectable at a cosmological distance 
if we perform a deep imaging (e.g. down to $i \sim 27.5$) within a few days. 
However, we also expect to find a few unrelated SNe within an error region of a FRB. 

We have compared the light curve and the color evolution 
of a kilonova predicted by the TH13 model 
to those of SN templates at various redshifts, 
and found that kilonovae can be distinguished from SNe 
by the color evolution [e.g. $\Delta(i-z)$] and/or the fractional variability ($\Delta f/f$). 
If real-time alerts of FRBs become available in near future, 
we suggest to perform deep imagings of positions of FRBs suggested 
to be at low redshifts (e.g. $z \lesssim 0.3$) from their DMs in red optical bands, 
starting within a couple of days and with intervals of several days. 
With an interval of 5 days and limiting magnitudes of $i = 27.5$, 
it would be possible to distinguish a kilonova 
with $i$-band magnitudes down to $i_{\rm 1st} \sim 27$. 

It should be noted that we have implicitly assumed 
that every NS-NS merger accompanies kilonova. 
Although a kilonova candidate is found associated 
with a short GRB 130603B \citep{Tanvir:2013a, Berger:2013a}, 
it is possible that NS-NS mergers which cause GRBs 
tend to eject more mass, producing brighter kilonovae. 
If other NS-NS mergers than the progenitor of GRB 130603B 
typically produce fainter (or no) kilonovae, 
we may find no optical counterpart even if FRBs originate from NS-NS mergers. 

The FRB rate is consistent with that of NS-NS mergers,
but at the high end of the plausible range \citep{Totani:2013a, Hassall:2013a}, 
suggesting that almost all NS-NS mergers produce FRBs. 
Kilonovae is an important candidate of the production site of $r$-process elements in the universe. 
Searches for kilonovae associated with FRBs may enable us 
to investigate what fraction of NS-NS mergers produce kilonovae, 
and to constrain the cosmic history of $r$-process element production. 

If the kilonova rate is close to the FRB rate reported by \citet{Thornton:2013a}, 
a blind search of kilonovae with a wide field optical camera is also interesting. 
When we take an image with a limiting magnitude of $i = 27.5$, 
similarly to the case of the FRB follow up observations discussed here, 
the expected number of kilonovae is $0.03\pm0.02$ per 1 deg$^{2}$, 
assuming the same FRB rate at $z < 0.3$ as in Section~\ref{sec:models}.

In the case that FRBs originates from WD-WD mergers, 
it is possible that they accompany SNe Ia as their optical counterparts, 
although the association of SNe Ia with WD-WD mergers remains a matter of debate. 
A SN Ia can be detected up to higher redshifts 
in the FRB follow up observations discussed above 
(up to $z \sim 0.9$ when observed 
down to $i_{\rm res}\sim27.5$ at 2 and 7 days after the explosion), 
and one can also search it around the expected peak time of the SN Ia 
(i.e. $\sim$ 20 days after the FRB in the rest frame). 

Since light curves of SNe Ia are well studied (e.g. \cite{Hsiao:2007a}), 
if a SN Ia is spectroscopically confirmed and several data points
of its light curve are obtained in an error region of a FRB,
it would be possible to constrain the explosion time with a day scale precision.
SNe with $i < 24.0$ can be spectroscopically classified using 8m-class telescopes, 
and the expected number density of unrelated SNe with $i < 24.0$ is 28.8 deg$^{-2}$ 
(i.e. 0.63 SNe within an error radius of 5 arcmin, figure~\ref{fig:ires_SN}). 
For SNe Ia, the magnitude limit of $i < 24.0$ corresponds to $z\lesssim0.6$. 
When we follow up a FRB whose DM corresponds to $z<0.6$, 
the expected rate of unrelated SNe Ia at $z < 0.6$ within an error radius of 5 arcmin 
is $< 0.003$ day$^{-1}$, assuming cosmic SN Ia rate density at $z < 0.6$ 
to be $< 5\times10^{-5}$yr$^{-1}$Mpc$^{-3}$ (e.g. \cite{Okumura:2014a}). 
Thus, if we find a SN Ia in an error region of a FRB 
with consistent explosion time, 
we would be able to conclude that the SN Ia is the counterpart 
of the FRB with a significant confidence level. 

\bigskip

We are grateful to the anonymous referee for helpful comments. 
Y.N. is supported by the Research Fellowship for Young Scientists 
from the Japan Society for the Promotion of Science (JSPS). 


\end{document}